%% file: FortnitePaper2.tex
\documentclass[journal]{IEEEtran}
\ifCLASSINFOpdf
\else
\fi

\usepackage[utf8]{inputenc}

\usepackage{graphicx}
\usepackage{multirow}
\usepackage{tikz}

\begin{document}
\title{How Players Play Games\\Observing the Influences of Game Mechanics}

\author{Philipp Moll, Veit Frick, Natascha Rauscher, Mathias Lux\\Klagenfurt University\\\{firstname\}.\{lastname\}@aau.at}% <-this % stops a space
\maketitle

\begin{abstract}
The popularity of computer games is remarkably high and is still growing every year. Despite this popularity and the economical importance of gaming, research in game design, or to be more precise, of game mechanics %-- game elements and rules defining how a game works -- 
that can be used to improve the enjoyment of a game, is still scarce. In this paper, we analyze \emph{Fortnite}, one of the currently most successful games, and observe how players play the game. We investigate what  makes playing the game enjoyable by analyzing video streams of experienced players from game streaming platforms and by conducting a user study with players who are new to the game. We formulate four hypotheses about how game mechanics influence the way players interact with the game and how it influences player enjoyment. We present differences in player behavior between experienced players and beginners and discuss how game mechanics could be used to improve the enjoyment for beginners. In addition, we describe our approach to analyze games without access to game-internal data by using a toolchain which automatically extracts game information from video streams. 
\end{abstract}


% \fi
\IEEEpeerreviewmaketitle

\input{content/intro}

\input{content/relatedWork}
\input{content/hypothesis}
\input{content/datacollection}
\input{content/dataanalysis}
\input{content/threatsToValidity}
\input{content/conclusion}
\bibliographystyle{IEEEtran}
\bibliography{bibliography}

% der sonstigen, nichtwissenschaftlichen Mitarbeiter, gibt

\begin{IEEEbiography}[{\includegraphics[width=1in,clip,keepaspectratio]{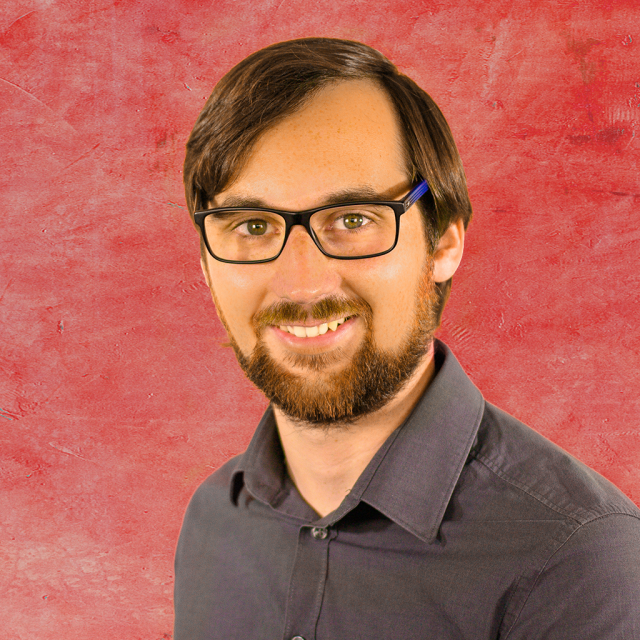}}]{Philipp Moll}
 is a PreDoc Scientist at the Institute of Information Technology (ITEC) at Klagenfurt University. He is currently pursuing a PhD in the field of computer networking. His main research interest is bringing information-centric networking architectures to online gaming, but he is also involved in work about reproducibility in research, multimedia communications and computer games. Philipp Moll received his MSc in computer science from Klagenfurt University in 2016.
\end{IEEEbiography}

\begin{IEEEbiography}[{\includegraphics[width=1in,clip,keepaspectratio]{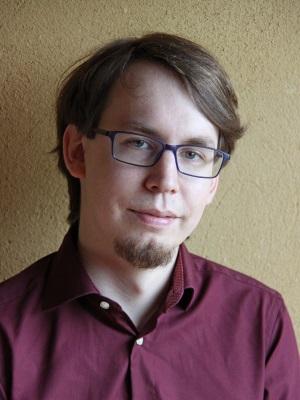}}]{Veit Frick} is a Senior Scientist at the Software Engineering Research Group (SERG) at Klagenfurt University. He is currently pursuing his PhD in software engineering. His research currently focuses on the detection of source code changes, their analysis and impact. He is involved in video game research and game studies. Veit Frick in one of the organizers of the Klujam, a biannual game jam that is a pillar of the local game development scene.
\end{IEEEbiography}

%\newpage

\begin{IEEEbiography}[{\includegraphics[width=1in,clip,keepaspectratio]{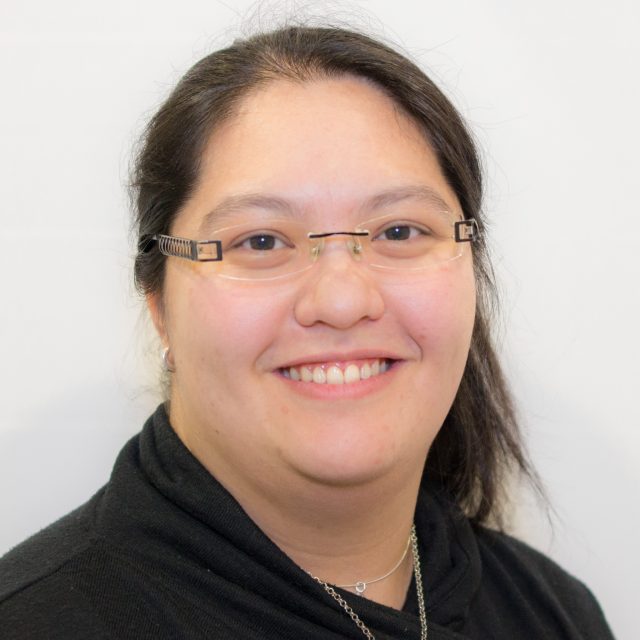}}]{Natascha Rauscher}
is a Student Research Assistant at the Institute of Information Technology (ITEC) at Klagenfurt University. She is currently pursuing a BSc in Applied Computer Science at Klagenfurt University. In the course of her work, she is analyzing the influence factors of the popularity of games on game streaming platforms and is co-organizing the biannual game jams held at Klagenfurt University.
\end{IEEEbiography}

\begin{IEEEbiography}[{\includegraphics[width=1in,clip,keepaspectratio]{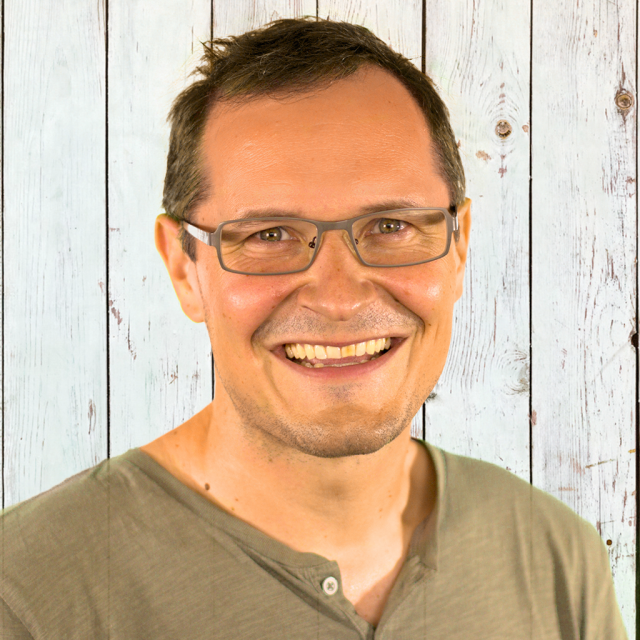}}]{Dr. Mathias Lux}
 is Associate Professor at the Institute of Information Technology (ITEC) at Klagenfurt University. He is working on user intentions in multimedia retrieval and production, semantics in social multimedia systems, and interactive multimedia in the domain of video games. In his scientific career he has (co-)~authored more than 100 scientific publications, serves in multiple program committees and as reviewer of international conferences, journals and magazines on a regular basis, and has (co-)~organized multiple scientific events. Mathias Lux is also well known for the development of the award winning and popular open source tools Caliph \& Emir and LIRE for multimedia information retrieval. He has integrated image indexing and retrieval features in the popular Apache Solr search server and his system is for instance powering the WIPO Global Brand Database. At Klagenfurt University he has established a lively community of game developers and enthusiasts who meet at regular events and game jams.
\end{IEEEbiography}


%\vfill


\end{document}

%% file: content/intro.tex
%Mögliche Konferenzen:

%\begin{itemize}
%	\item IEEE CIG - März
%	\item DIGRA - Feb 5
%	\item IEEE GEM - April
%	\item ACM CIE - Cont.
%	\item ACM MMVE - März
%   \item PeerJ Computer Science - Cont.
%\end{itemize}

%Eventuell auch interessant: Wie verhalten sich Spieler jetzt im Vergleich zu vergangenen Update.

\section{Introduction}

% Observed two things, beginners and experienced players. Why do we do that?
% Es gibt einen Unterschied zwischen dem Verhalten von Anfänger und Fortgeschrittene. Wir wollen den Zeigen. Andere Game Designer können dann davon lernen und ihre spiele anfängerfreundlich gestalten.
% Was ist Game Mechanic - Zitat Paper [sicart]

Gaming has become a fundamental part of the entertainment industry. In 2017 the US video game industry alone generated USD 36 billion in revenue and 2.6 billion people worldwide played video games~\cite{esareport}. The key feature that sets games aside from other forms of media is interactivity. %People do not only perceive games, they play and interact with them. 
This interaction is shaped by a fixed set of rules, also known as game mechanics. Understanding the interactions between players and the game mechanics of a game can help to improve game design to further increase the players' satisfaction and enjoyment.

We propose an approach to research how players play games by investigating their interaction with the game mechanics using the free-to-play game \emph{Fortnite}. A main obstacle when researching the influence of game mechanics on player interaction is gathering data. As with most other software, it is possible for game developers to log information on how players interact with a game. However, in most cases such data is not released to the public due to commercial and data protection reasons. For games, however, there is another way to gather such data. There are countless video streams of gaming sessions for most popular games freely available online. We describe how to extract information about player interaction from such video streams. Additionally, we conducted a user study with 13 beginner Fortnite players and compare their behavior in the game to that of experienced Fortnite streamers. %By analyzing the differences between beginners and more experienced players we can point out what elements of game mechanics influence enjoyment and satisfaction of beginner players and how they can be used to hook novel players.
We argue that our findings demonstrate that by analyzing the differences between beginners and experienced players, it is possible to identify the elements of game mechanics which influence enjoyment and satisfaction among beginner players. We also discuss how these elements of game mechanics can be used to motivate further gaming.

\section{Fortnite Battle Royale}

We use the game mechanics of the \emph{battle royale} game mode of the game Fortnite by Epic Games for this study. Fortnite has been released in 2017 and its free-to-play battle royale game mode has quickly grown to be one of the most played games worldwide \footnote{https://www.epicgames.com/fortnite/en-US/news/fall-skirmish-details, last accessed: 2018-10-15}. %Almost 80 million players played Fortnite in August 2018 alone\footnote{https://www.epicgames.com/fortnite/en-US/news/fall-skirmish-details, last accessed: 2018-10-15} and 125 million player have played it since release\footnote{https://www.epicgames.com/fortnite/en-US/news/announcing-2018-2019-fortnite-competitive-season, last accessed: 2018-10-15}.
The large number of players allowed us to easily find video streams in sufficient quality to analyze.

In Fortnite's competitive online mode battle royale, 100 players are fighting against each other and the last player to survive wins the game. 
Fortnite is split in several phases. During the first phase, the \textit{lobby}, people join the game. Once enough players have joined, all players fly together in a flying bus across the map of Fortnite which is one single large island. In this phase, called \textit{jump phase}, players can choose when to parachute down to the island and where to land. Once landed, players can start to fight other players and collect loot, such as weapons, healing bandages, shield potions, and other useful items to help them survive and kill other players. After a fixed amount of time, a storm appears on the island and all players who are not in the eye of the storm continuously lose health points. The two phases \textit{storm brewing} and \textit{contraction} alternate until the end of the game. 
%During these two phases the eye of the storm is getting smaller and smaller.
During the storm brewing phase players are shown which part of the island is going to be safe in the future. In the contraction phase the eye of the storm is getting smaller and smaller. This mechanic shrinks the area of the game over time and compensates for the smaller number of players that are still alive later in the game.

%% file: content/relatedWork.tex
\section{Related Work}
\label{sec:relatedWork}

Analyzing and evaluating games is not a new concept.
A definition of game mechanics is given by Sicart~\cite{definingGameMechanics}.
Sánchez et al.~\cite{playability} present the concept of \textit{playability} as a framework for the analysis of user experience in video games. Sweetser et al.~\cite{Sweetser:2005:GME:1077246.1077253} describe \textit{GameFlow}, a model to evaluate player enjoyment. Pinelle et al.~\cite{Pinelle:2008:HEG:1357054.1357282} present a heuristic for detecting usability problems in games.
These approaches focus on the user while in our work we present a way to analyze how game mechanics are used and what their influences are.

%The related question on why we play videogames
%http://twvideo01.ubm-us.net/o1/vault/gdc04/slides/why_we_play_games.pdf Why we play games
%McGonigal
%Psychoteil Masterarbeit (Flow etc)

%Game mechanics themselves have also been subject to analysis in the field of game studies. Sicart~\cite{definingGameMechanics} discusses a definition of game mechanics and Järvinen~\cite{book} gives further insight in how they function.

Deterding et al.~\cite{Deterding:2011:GDE:2181037.2181040} define \textit{gamification} as the use of game design elements (game mechanics) in non-game contexts. The influence of gamification has been researched in multiple studies.
%The influence of game mechanics, on non-game systems (gamification) has already been researched by multiple studies.
Montola et al.~\cite{montola_nummenmaa_lucero_boberg_korhonen_2009} analyze the effect of applying achievements to a photo sharing service and Anderson et al.~\cite{anderson_huttenlocher_kleinberg_leskovec_2013} analyze how adding badges to websites can steer user behavior. Multiple studies investigate the influence of gamification in education, Dicheva et al. provide an overview.

Hamari et al.~\cite{Hamari2009GameDA} analyzes how game mechanics can be used to create demand for virtual goods. Additionally, the game mechanics of Fortnite have been used to analyze and simulate the network traffic of games \cite{NetworkTraffic}.

%Probably not to use:
%Federoff "HEURISTICS AND USABILITY GUIDELINES FOR THE CREATION AND EVALUATION OF FUN IN VIDEO GAMES" - how games are created

%As mentioned in the introduction, the authors are not aware of research analyzing game mechanics themselves in an empirical way. However games, and their core element the game mechanics have been analyzed from a multitude of different fields. Game mechanics have also been subject to further analysis in the field of game studies \cite{definingGameMechanics} \cite{book}

%Anderson provides an overview of studies regarding the effects of playing violent video games \cite{anderson_2004}. 

%Maybe adept this to fit the toolbox how to analyze games more closely

%% file: content/hypothesis.tex
\section{Hypotheses}
\label{sec:hypothesis}

%\begin{itemize}
%	\item State our Hypothesis + possible impact both
%	\item Write about how they are backed up by the data (or not) both
%	\item Write about limitations of our approach. both
%\end{itemize}

In this section we propose four hypotheses and use them as basis for our investigation on how players interact with games and discuss possible impacts. Our hypotheses help to show differences in behavior of experienced and beginner Fortnite players. In addition, the influence of game mechanics on the players' satisfaction with their game rounds as well as their enjoyment is investigated. 

%To find out how players interact with Fortnite, or more general with games, and to find out if beginners behave differently than experienced players, we formulated four hypotheses, which are examined and discussed in the rest of the paper. In the following, our hypotheses including a discussion of possible impacts are presented.

%\begin{description}
\textbf{H1} \emph{Satisfaction and enjoyment are correlated and are influenced by the player's success in the game.} 
Playing a new game for the first time can be complicated and challenging. We suspect that the amount of enjoyment a beginner experiences and their satisfaction with their own play are influenced by their success in the game. If this hypothesis holds true, game designers can influence the enjoyment and satisfaction of beginners by guiding them towards more successful games.

%Wenn true, dann können wir davon ableiten dass es wichtig ist beginner zum success zu führen.

\textbf{H2} \emph{There are frequently visited locations with only low killing activity.} 
One of the main activities in Fortnite is to fight against other players. Due to the characteristics of the map, we suspect that there are places, which are visited by many players, but where only a few confrontations take place. We call those places boring spots and assume that their existence influences the gaming experience. Game designers who are aware of boring spots can remedy or promote them by changing the characteristics of the map.

%Wenn true, can be used to improve the map in the future. by changing map...

\textbf{H3a} \emph{Experienced players choose different landing spots than inexperienced players.} 
We assume that experienced players, who have already played Fortnite and thereby implicitly have a better knowledge of the game mechanics, play differently compared to beginners. We observe landing spots of beginners and compare them to the landing spots of experienced players to validate this assumption.

\textbf{H3b} \emph{The enjoyment of Fortnite beginners is influenced by the chosen landing spot.} 
We suspect that the landing spot influences the course of a game and thereby the player's enjoyment. If both parts of this hypothesis hold true, the game mechanics can be used to steer the behavior of beginners and experienced players separately. In addition, this supervision can be used to increase the players' enjoyment.

%Wenn beide true und positive korrelation, dann gut, ansonsten müssen anfänger besser durch spiel geführt werden. sollte von game designern beachtet werden.

\textbf{H4} \emph{The amount of time playing other games influences the success when starting to play Fortnite.} 
Although game mechanics of different games, especially if they belong to different genres, can differ significantly, we assume that there is a correlation between the regularity of playing other games and the success when starting to play Fortnite. Experience in playing games teaches common patterns in game design and enhances familiarity with conventional control schemes, which we suspect helps in performing well at Fortnite.

%\begin{enumerate}
%	\item H1: The landing spot (LS) influences the success of a player. (Success is different from enjoyment (kills, place))
%	\item H2:The LS of experienced players differs from the LS of inexperienced players.
%	\item H3: There are LS hotspots.
%	\item H4: There are killer/death/killing hotspots. (killing = killer + death)
%	\item H5: There are activity hotspots which are no killing hotspots. (aka boring points)
%	\item H6: Satisfaction and enjoyment are correlated (Questionnaire)
%	\item H7: Enjoyment is influenced by whether the LS is in a LS hotspot or not for inexperienced players.
%	\item H8: Enjoyment is influenced by whether the player frequents killing hotspots or not for inexperienced players.
%	\item H9: A person who plays games regularly is good at playing Fortnite.
%\end{enumerate}

%% file: content/datacollection.tex
\section{Gathering Gameplay Information}
\label{sec:approach}
%\begin{itemize}
%	\item Explain how we extract data from streams. philipp
%	\item Explain how we set up the experiment, the questionnaire, the participants. veit
%	\item Write about what data we have with the videos and questionnaire + why we collected/selected them + data quality (no inexperienced players in the streams) both
%\end{itemize}

%We collected data from Fortnite beginners as well as data from experienced Fortnite players in order to test the presented hypotheses. Gathering data of Fortnite beginners is done by conducting a user study, which is explained in detail in Section \ref{sec:userstudy}. Data from experienced players is collected by analyzing game video streams from the game streaming platform \emph{mixer}\footnote{https://mixer.com, last accessed: 2018-09-27}. Details on the selection of players, the process of analyzing videos and the evaluation of the produced data is given in Section \ref{sec:streamer}.

\begin{figure}[ht]
	\centering
	\includegraphics[width=0.9\columnwidth]{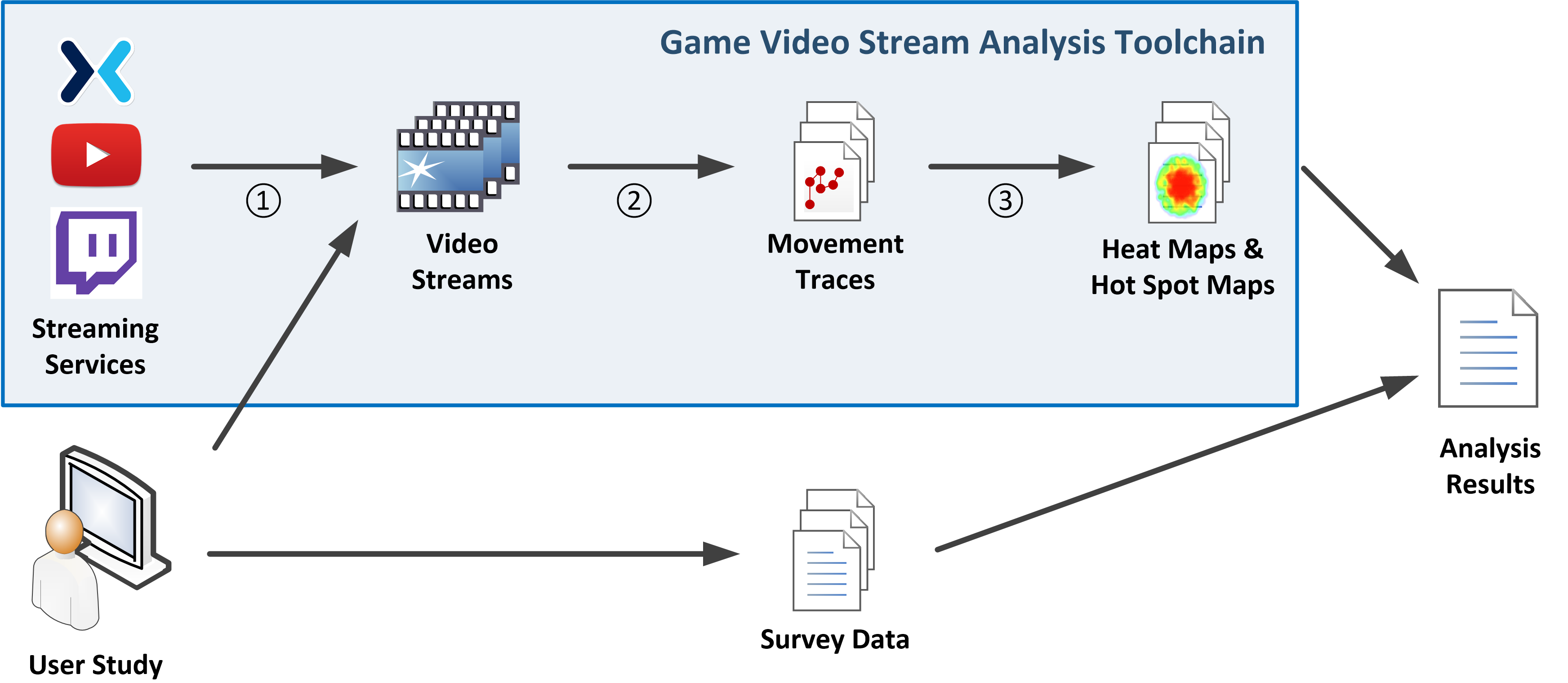}
	\caption{Visualization of our evaluation process. The \textit{game video stream analysis toolchain} is highlighted in blue. The individual steps of the toolchain (video downloading, information extraction, and information visualization) are labeled by circled numbers.}
	\label{fig:toolchain}
\end{figure}

Testing our hypothesis required collecting data of Fortnite beginners as well as experienced players. The process used to collect and analyze data is based on our \textit{game video stream analysis toolchain} and is illustrated in Figure~\ref{fig:toolchain}.

We first conducted a user study with beginner Fortnite players resulting in survey data and game video streams, which is explained in Section~\ref{sec:userstudy}. We then collected data of experienced players by analyzing game streaming footage from the game streaming platform  \emph{mixer}\footnote{https://mixer.com, last accessed: 2018-09-27}. We used our toolchain to extract and visualize data of individual game rounds for later analysis. More information about the collection of data of experienced players and about the game video stream analysis toolchain is described in Section~\ref{sec:streamer}.

\subsection{Collecting Data of Beginners}
\label{sec:userstudy}

To collect the data of beginner Fortnite players, we conducted a user study with a total of 12 participants. Participants did not have to meet any specific requirements but had to have only a limited experience (less than 4 hours) in playing Fortnite on any platform.

For the user study we first explained the process of the study to the participant and asked for their consent on recording their gaming session, including the audio captured from their headset microphones. All gathered data was anonymized. We then asked the participant to fill out a questionnaire consisting of 13 questions concerning demographic background, general gaming experience (platform, genre, and playtime), experience in watching video streams of games, and experience in playing and watching Fortnite specifically. %The full questionnaire can be found in the Appendix.
Afterwards we asked the participant to play as many rounds of Fortnite as they like. We provided the hardware for the players (PS4 Pro, standard controller, headset, HDTV) and gave a brief explanation on how Fortnite works. We asked the participants to verbalize their thoughts while playing. Some participants asked questions during their sessions. While we did answer questions concerning how to correctly interact with the game (e.g. about how the player avatar is controlled), we did not answer questions about tactics nor did we offer any kind of advice.

We recorded their gaming sessions by live-streaming the session, including the audio from their headset microphones, to a private YouTube channel. The video footage uploaded to YouTube has been analyzed as described in Section~\ref{sec:streamer}. Due to technical reasons, we were not able to use all footage, but the data from 87 out of 101 games was analyzed.

After each game, we asked the participants to rate their satisfaction with the game round regarding their personal expectations, their enjoyment during playing the round, and to state their strategy for the round, if they had any. After their last game, we asked the participants to rate how comfortable they felt during the user study and their overall satisfaction and enjoyment. Additionally, we asked what elements of Fortnite made the game interesting for them and what elements did make it less interesting.

%TODO Add why we collected this data.

\subsection{Collecting Data of Experienced Player}
\label{sec:streamer}

A number of reasons, e.g. the local unavailability of experienced Fortnite players, makes it unreasonably hard to conduct a user study to collect data of experienced players. Furthermore, to test our hypotheses, a great diversity of players as well as a large number of played games are required. Instead of observing the behavior of experienced players in a user study, we decided to examine the large amount of videos available on game streaming platforms. %We examine recorded gaming videos of streamers. A streamer is a player, who plays games and streams their game session live to game streaming platforms. On those platforms, the streamer has the possibility to interact with their audience by reading chat messages and verbally responding to them. These live game streams are currently very popular and reach thousands of viewers. %maybe a quote?

%We decided to use the large amount of videos available on streaming platforms to analyze experienced players' playing behaviour.

%\subsubsection{Selection of Streamers and Videos}

\noindent
\textbf{Selection of Streamers and Videos: }
%In addition to the gaming behavior of beginners, which was collected in the course of the user study, the behavior of experienced players needs to be observed as well.
The selection of streamers is critical for several reasons. First, if the streamer is playing Fortnite for the first time, they can not be classified as experienced player and might lack the knowledge and tactics of an experienced player. Second, to automatize the analysis of the experienced players' behavior, the videos have to be in a uniform format. This means that video properties, such as the resolution, have to be standardized. Additionally, game specific elements, such as the head-up display (HUD), need to be displayed on predefined positions on the screen. To get only data of experienced players, we focused on players who have streamed Fortnite frequently and provide several hours of Fortnite streams on their streaming profiles. We assume that players who frequently stream themselves playing Fortnite are experienced Fortnite players. %Get rid of the professional - meaning unclear % Done

Fulfilling the second requirement, uniform format of gaming videos, is more challenging. First, the position of the HUD, which contains important information about the game, such as the number of kills and the number of remaining players, is dependent on the used resolution and aspect ratio. Furthermore, diverse in-game overlays, such as displayed buttons, differ on different gaming platforms\footnote{Fortnite frequently displays buttons to notify the player of possible actions, e.g. the possibility to open a chest or to pick up an item. The style of the displayed buttons varies on different platforms.}. %Maybe not in  a footnote?
In addition, the automated analysis of gaming videos requires a resolution of at least 1080p. Video encoding artifacts in lower resolutions prevent the reliable execution of certain tasks, such as optical character recognition (OCR) and minimap matching. Limited by these technical details, we restricted our analysis of game streams to 1080p videos resulting from a single hardware platform. For technical reasons, we decided to choose the Xbox One family as platform. % VF All Xboxes? PM the whole Xbox One familiy should now include One S and One X - VF Why not PS4? %PM For technical reasons should do the job.
Based on these restrictions, we selected videos from eleven different streamers from the streaming platform mixer.
%For downloading the videos we used \emph{youtube-dl}\footnote{https://github.com/rg3/youtube-dl, last accessed: 2018-12-04} -- representing the first tool of our toolchain -- which allows to download video streams from a broad range of websites.
In total, we downloaded 234~hours of game streams, including almost 900 unique games. 

%\subsubsection{Extraction of Information}
%\label{subsubsec:extracting}
\noindent
\textbf{Extraction of Information: }
%The selection of game video streams is an important step, but results in video files only. 
Analyzing the player's behavior requires more tangible information than just video files. This could be information about how players move during a game, spots where they begin the game, or spots where people die. To extract this information, tracing the players' positions and actions is required. This analysis represents the second step of our toolchain and includes a combination of computer vision tasks, which are explained in more detail in the following.

\begin{figure}[ht]
\centering
	\includegraphics[width=0.9\columnwidth]{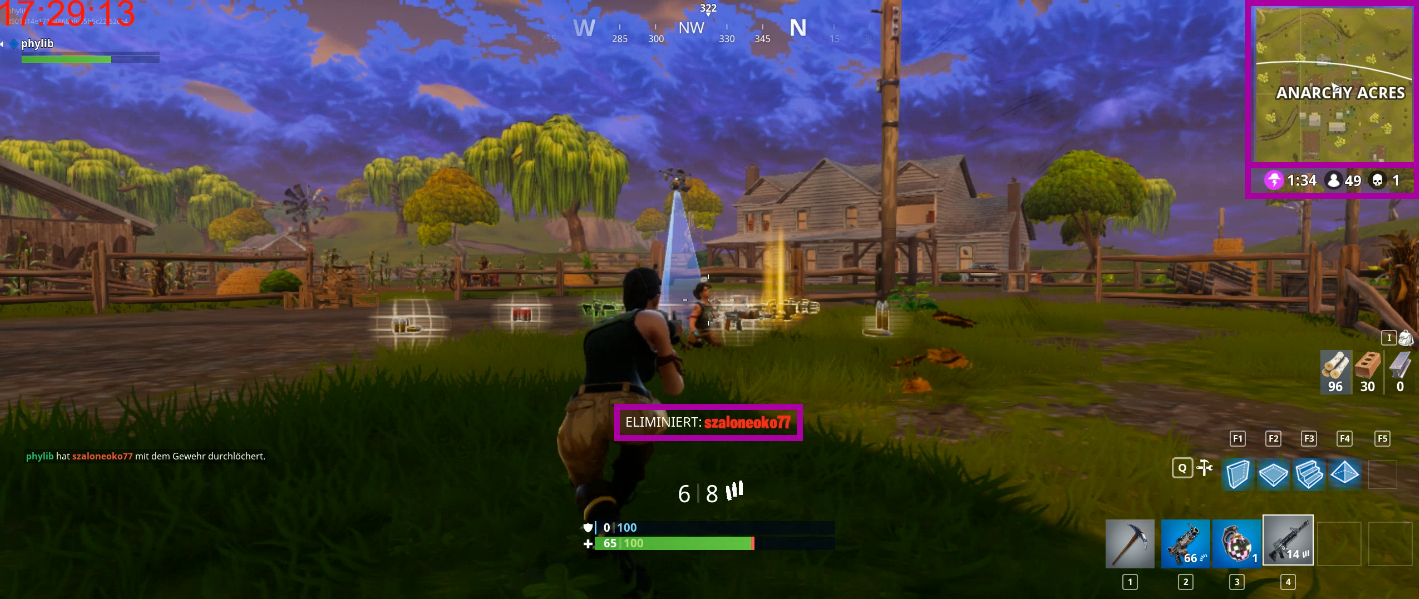}
	\caption{Screenshot of playing Fortnite, with examples of used game information highlighted in purple.}
	\label{fig:screenshot}
\end{figure}

The main information visible in Fortnite videos is highlighted by purple boxes in Figure~\ref{fig:screenshot}. The minimap in the upper right corner shows a small portion of the overall map and visualizes the current position of the player. The player's position is denoted by a small triangle in the center of the minimap. 
Below the minimap, information about the current game is displayed. This information includes the current phase of the game, the current number of active players, and the current player's number of kills.

Besides, the data which is always displayed during the game, additional information about certain events occasionally pops up on the screen. Figure~\ref{fig:screenshot} shows information about a recent kill, highlighted in the center of the screen.

The selected videos mostly contained several consecutive game rounds. Accordingly, we first identified the start and end points of each single game in each video by tracing the game's phases. For this we tracked the phase icon below the minimap. %The begin of a game is identified by the \textit{jump phase}. % which is denoted by the third icon in Figure~\ref{fig:gamePhases}.
%After the jump phase, several \textit{storm brewing} and \textit{contraction} phases alternate until the game ends, either by winning the game or by getting killed. A game that the player has won can be identified by a check icon, the death of a player is identified by either a pop-up information on the screen, or by a slight movement of the HUD.
%\begin{figure}[ht]
%	\centering
%	\includegraphics[width=0.5\columnwidth]{images/phases}
%	\caption{Icons representing the phases of a game. \textit{The lobby} is the first phase, where up to 100 players are assigned to the game. \textit{The jump phase}, where players decide on their starting position can be seen as the start of the game.}
%	\label{fig:gamePhases}
%\end{figure} 
After identifying the begin and end times of each game in a video, the player's movements are traced by analyzing the minimap. The small portion of the map shown by the minimap is matched to the full image of the map by using OpenCV's template matching algorithm. To improve the accuracy and to minimize the consequences of overlays, images are preprocessed by using local contrast enhancement.
Finally, the OCR tool Tesseract is used to extract the number of active players and the number of kills of a player, which are stored together with the player's position for later analysis.

%VF TODO Maybe add a small overview over all the data that can be captured this way
%PM Done
A summary of the data extracted from Fortnite videos is shown in Table~\ref{tab:extractedData}. Data which was extracted directly from the screen is referred to as \emph{raw data} and is sampled once per second. Based on this data, additional information about the game, such as the final place of a player can be calculated. Values marked with an asterisk ($^*$) were extracted and can be used for analysis, but are not used to answer our hypotheses.

\begin{table}[h]
	\centering
	\small
	\begin{tabular}{l|l}
		\textbf{Raw Data} & \textbf{Calculated Values}    \\
		\hline\hline
		\multirow{4}{*}{Player position} & Landing spot \\
		 & Movement trace \\
		 & Death position (If player died) $^*$ \\
		 & Win position (If player won) $^*$  \\
		\hline
		\multirow{2}{*}{No. of active players} & Final place \\
		 & Active player progression $^*$ \\
		\hline
		\multirow{2}{*}{Game phase} & Start/End of game                \\
		 & Duration of game phases $^*$           \\
		\hline
		Kills & Total number of kills                             \\
		\hline
		Kills + player position & Killing positions        
	\end{tabular}
	\caption{Overview over data extracted from Fortnite streams.}
	\label{tab:extractedData}
\end{table}

%\subsubsection{Using the Extracted Information}
%\label{sec:usingInformation}
\noindent
\textbf{Using the Extracted Information: }
After collecting movement traces and other relevant data of each game, the information needs to be evaluated and interpreted, which is done by using \textit{heat maps}. For each analysis, a set of relevant games is selected (e.g. all games played by Fortnite beginners) and the movement traces of those games are visualized in a heat map, where places with high activity (players visiting the place) are colored with more intense colors than places with low activity. In addition to activity heat maps, heat maps for landing spots and for spots where players are killed are created.

Based on these heat maps, \textit{hot spot maps} can be extracted. A hot spot is defined as region, where the activity was above a certain threshold. A hot spot map is created by binarizing the heat map and by reducing to small spots by eroding and dilating the resulting image. Based on these heat maps and hot spot maps, the presented hypotheses can be examined.

%% file: content/dataanalysis.tex
\section{Influence of Game Mechanics}
\label{sec:analysis}

In this section we use the presented tools to evaluate the hypotheses presented in Section~\ref{sec:hypothesis} and describe our findings. \footnote{All correlations in this section are calculated with Spearman's rank correlation coefficient. Calculations are available in the reproduction set.} 

\textbf{H1:} Based on the survey of the user study with Fortnite beginners, we want to find out whether \emph{satisfaction} and \emph{enjoyment} are correlated for beginners. During the user study, players rated their satisfaction and their enjoyment of each individual game round on a five-level Likert scale. A first analysis of the answers showed a medium correlation between satisfaction and enjoyment ($r_s(85)=0.47,~\rho<0.001$). This means that beginners tend to enjoy games when they are satisfied with their own performance.

Furthermore, we found a strong correlation ($r_s(85)=0.61,~\rho<0.001$) between satisfaction and the duration of a game. %VF TODO need to reformulate. We did NOT check for independence. Might still be the case! %PM Done
In addition, beginners are more satisfied with a game, when they achieve a better (lower) place ($r_s(85)=-0.53,~\rho<0.001$) and when they achieve more kills ($r_s(85)=0.48,~\rho<0.001$). Similarly, beginners' enjoyment depend on parameters like kills, place and length of the game, but here only weak correlations are observed.

These correlations allow to confirm our initial assumption that \emph{satisfaction and enjoyment are correlated.} In addition, we saw a strong correlation between the players' success in terms of the duration a player survives and their satisfaction with the game.

\begin{figure}[ht]
	\centering
		\includegraphics[width=0.6\columnwidth]{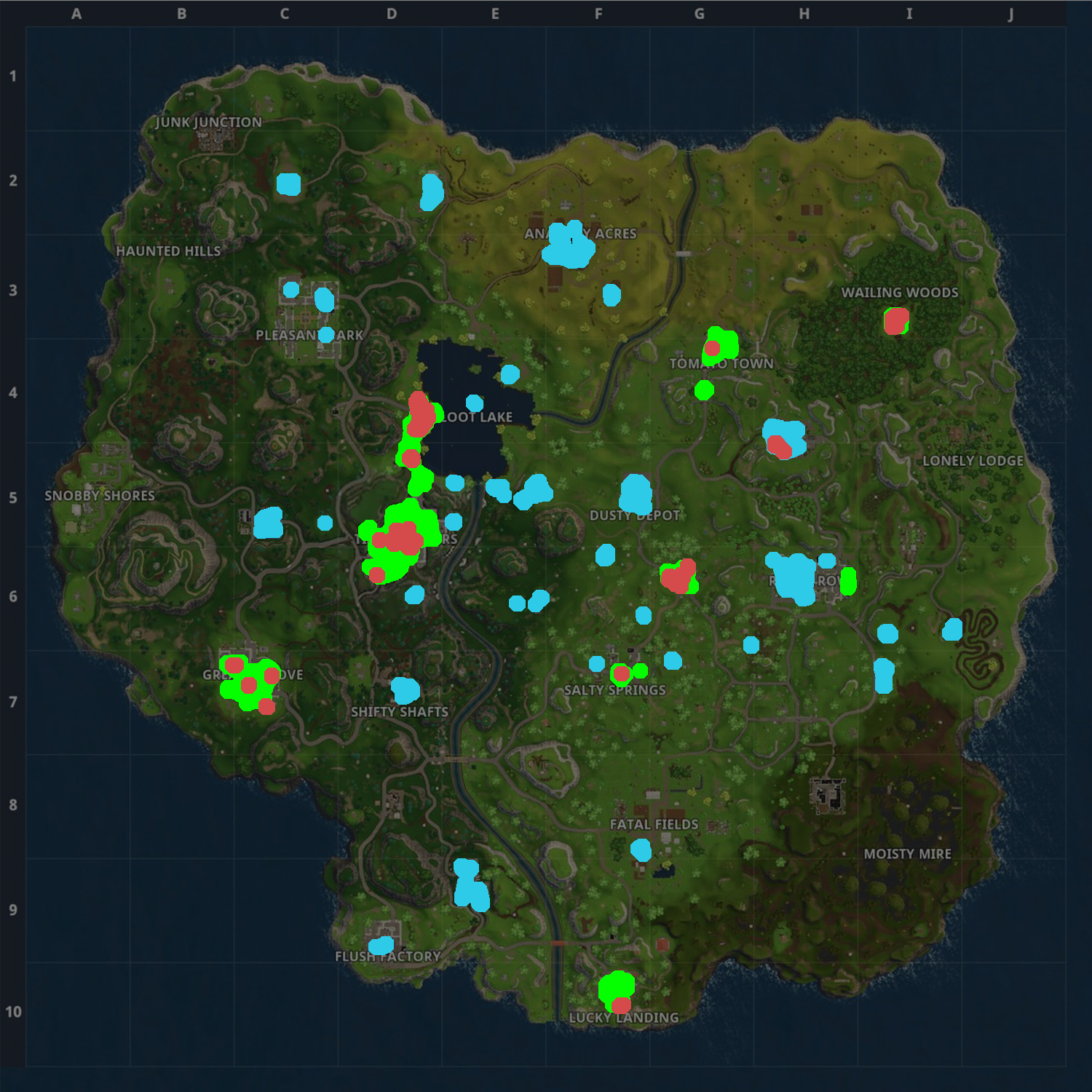}
	\caption{Killing hot spots are highlighted in red. Activity hot spots that include killing hot spots are highlighted in green, activity hot spots without high killing activity in blue.}
	\label{fig:h5}
\end{figure}

\textbf{H2:} Predetermined by the map and structures on it, as well as its geography, certain areas of the map are frequented more often than others. 
For example, in \emph{Tilted Towers}, a small city with many skyscrapers, a lot of weapons and other loot that can be used by the player later on in the game is hidden.
We refer to those more attractive areas as \emph{activity hot spots}. Similarly, there are spots where a lot of fights occur and thereby a lot of players are killed, which we refer to as \emph{killing hot spots}. Figure~\ref{fig:h5} highlights killing hot spots in red and activity hot spots in blue and green. Blue areas denote activity hot spots with comparatively low killing activity, in green areas a lot of fighting and killing takes place. As visible, hot spots without killing activity exist, which supports our hypothesis of so-called \emph{boring spots}. A boring spot is defined as a location which is visited by a lot of players, but where no, or almost no, fights occur.

\begin{figure}[ht]
	\centering
	\begin{minipage}{0.32\columnwidth}
		\includegraphics[width=\textwidth]{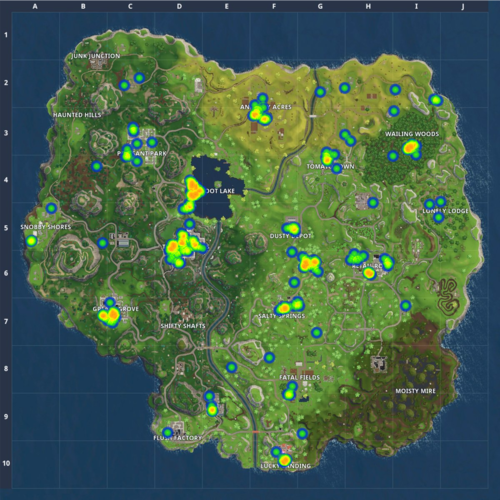}
	\end{minipage}
	\begin{minipage}{0.32\columnwidth}
		\includegraphics[width=\textwidth]{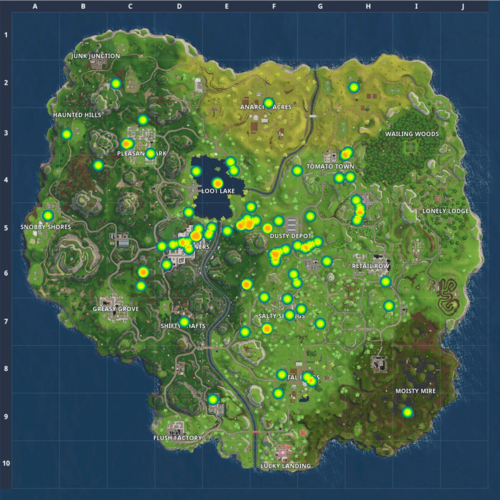}
	\end{minipage}
	\begin{minipage}{0.32\columnwidth}
		\includegraphics[width=\textwidth]{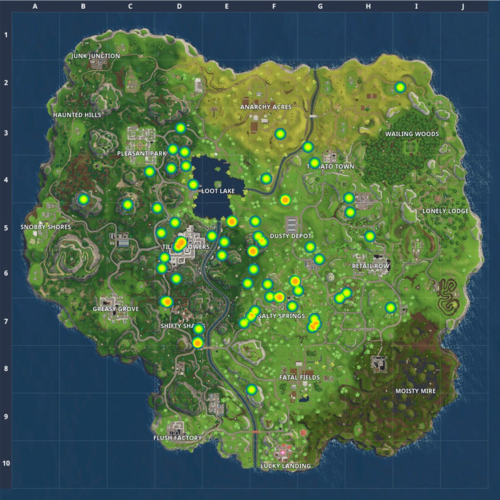}
	\end{minipage}
	\caption{Changes of the killing heat map over time. The left heat map shows game minutes 3 and 4, the center heat map  minutes 6 and 7 and the right heat map minutes 12 and 13.}
	\label{fig:killsovertime}
\end{figure}

In addition, we could observe that hot spots change over time. In Figure~\ref{fig:killsovertime} the killing heat maps at three different stages of the game are visualized. At the beginning of the game (left heat map), most kills take place on activity hot spots. The later in the game (center and right heat map) the more distributed the kills get. This is primarily due to the shrinking eye of the storm, in which players do not get damage. A Fortnite game typically lasts between 20 and 25~minutes. After only 12~minutes, almost no clustering around activity hot spots can be observed. 

\begin{figure}[ht]
	\centering
	\begin{minipage}{0.49\columnwidth}
		\includegraphics[width=\textwidth]{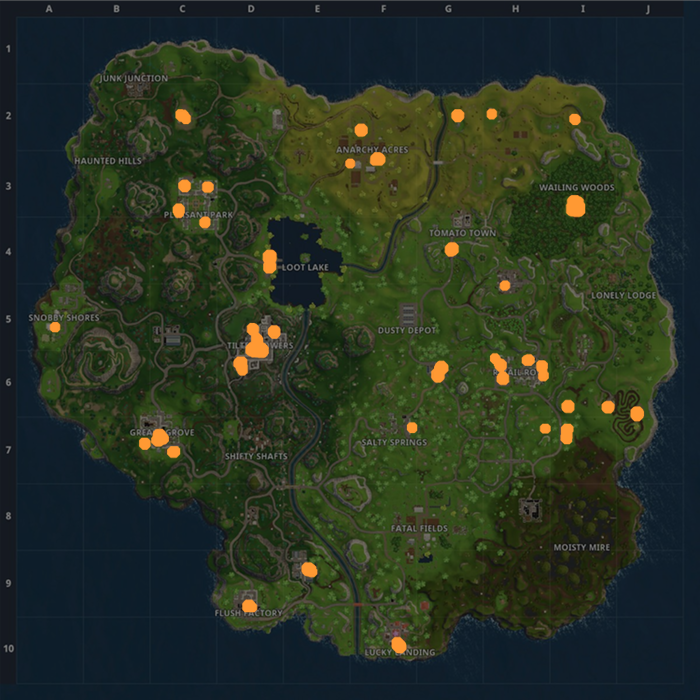}
	\end{minipage}
	\begin{minipage}{0.49\columnwidth}
		\includegraphics[width=\textwidth]{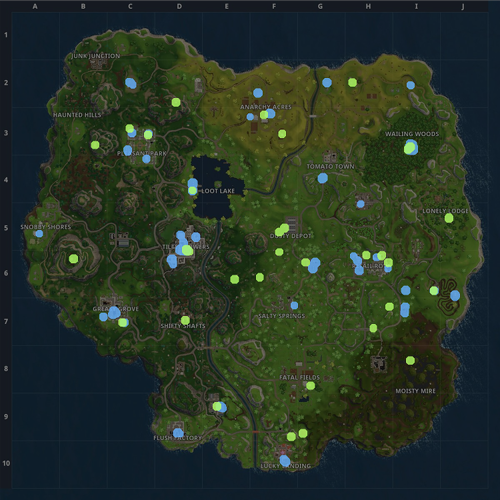}
	\end{minipage}
	\caption{Hot spot map showing landing hot spots for all players (orange) on the left and separated for experienced (blue) and inexperienced (green) players on the right.}
	\label{fig:h2}
\end{figure}

\textbf{H3a:} %In Fortnite, specific locations on the map attract the attention of many players. For instance, \emph{Tilted Towers}, a small city with skyscrapers is well-known for large fights and loads of loot for the players to pick up.
Similarly to \emph{H2}, we suspect that there are spots which are more popular than others for starting the game. By extracting the landing spots, which basically are places where players choose to start the game, a landing hot spot map can be created. % as explained in Section~\ref{sec:streamer}. 
The landing hot spot map in Figure~\ref{fig:h2} (left part) shows the landing hot spots over all observed games and reveals that Tilted Towers and other well-known locations strongly impact the landing behavior of players.

%When comparing the landing hot spots of experienced players with those of inexperienced players, we can see major differences.
We can see major differences between the landing hot spots of experienced players and the landing hot spots of inexperienced players. The right part of Figure~\ref{fig:h2} visualizes the landing hot spots of experienced (blue spots) and inexperienced players (green spots). Some exceptions aside, experienced players mostly follow one of two strategies. They either land on the edges of the island, or directly on one of the activity hot spots. We assume that landing on the edges of the island is a good possibility to collect weapons or other loot without facing an early confrontation. Landing on an activity hot spot, however, most likely leads to lots of action and possibly many kills in an early stage of the game and great loot when the initial action is survived.

When focusing on inexperienced players we can see that landing spots appear more random than those of experienced players. This difference in landing behavior confirms \emph{H3a} -- \emph{Experienced players choose different landing spots than inexperienced players}.

\textbf{H3b:} Now that we know that landing hot spots, activity hot spots and killing hot spots exist, we want to analyze if landing in those hot spots influences the enjoyment of Fortnite beginners. We analyzed the game rounds played in the user study and checked if the players landed in landing hot spots.%, in activity hot spots or in killing hot spots. 

We observed a weak negative correlation between landing on landing hot spots and enjoyment ($r_s(85)=-0.27,~\rho=0.01$) for beginner Fortnite players. Furthermore, the analysis also shows weak correlations between landing in an activity hot spot and the achieved place ($r_s(85)=0.31,~\rho=0.004$) and landing in an activity hot spot and the game duration ($r_s(85)=-0.31,~\rho=0.004$). The achieved place, as well as the game duration, both correlate with enjoyment. %These correlations indicate that beginners achieve a worse place  when they land on a spot with a lot of activity, resulting in lower enjoyment.

In summary, we observe that beginners landing away from both, landing and activity hot spots tend to have  a more enjoyable experience. This supports our hypothesis that \emph{the enjoyment of Fortnite beginners is influenced by the landing spot.}

\textbf{H4:} As stated in the introduction, Fortnite belongs to the battle royale genre, where the last surviver wins. Nevertheless, one of the main strategies to achieve a good result is killing other players. This is why we assume that Fortnite beginners who regularly play other games, especially games from the shooter genre, perform better in Fortnite. Although, there is a strong correlation between achieving a good place and the number of kills in a game ($r_s(811)=-0.63,~\rho<0.001$), we found no correlation between achieving a good place and regularly playing other video games for Fortnite beginners. Also, we found no correlation between the amount of time spent playing other games and the success in Fortnite (achieved place, number of kills, and game duration).

Interestingly, for beginners we found a weak correlation between the amount of watched Fortnite streams and the number of times a hot spot was chosen as landing spot ($r_s(85)=-0.37,~\rho<0.001$). This indicates that players, who already know about the game mechanics from watching streams, are able to use their knowledge. However, no indication for improved success or satisfaction was found.
When interpreting these results, we can not confirm our hypothesis that \emph{the amount of time playing other games influences the success when starting to play Fortnite}.

%% file: content/threatsToValidity.tex
\section{Limitations}
\label{sec:threats}

%The participants for the user study have been recruited from the authors' social circles and are therefore not necessarily representative. Additionally, 

For the user study with Fortnite beginners, no female participants registered. Therefore, we do not know if the results also apply for female Fortnite beginners.

%Phan et al. describe that video game preferences of men and women differ [https://journals.sagepub.com/doi/10.1177/1071181312561297], however, we are not aware of any research indicating differences in their respective playstyles.  

%However, we otherwise managed to recruit a diverse group of participants also in terms of general gaming experience.

%Nevertheless, our results do apply for male Fortnite players.

%We do not believe that gender or the social circle have influence on our findings.

A second limitation to validity is that the user study setting did not necessarily resemble the normal gaming environment of the players since the participants used our provided hardware and were asked to verbalize their thoughts. Nevertheless, 11 out of 12 participants felt comfortable or very comfortable during the user study.

Fortnite is an actively developed game and new patches are constantly published. It is, therefore, not possible to repeat the user study at our patch level. However, the analysis of the questionnaires and our videos is reproducible. Future studies, gathering data from different Fortnite patch levels, might yield other results. A comparison might therefore be interesting and could provide additional insight.

Our toolchain is tailored towards analysing Fortnite videos and uses the characterisitcs of the Fortnite HUD. Even though we think that our toolchain can be adapted to work with other games, some games might have too little recognizable features in their HUD to be properly analysed.

%Due to technical reasons, we gathered our data for the experienced players from the mixer streaming service and only used Xbox footages for analysis. Data from other streaming services or platforms might differ. However, the mechanics of Fortnite are the same on all platforms and it even allows for cross-platform play. Therefore, we do not believe that our results are influenced by  this.

%% file: content/conclusion.tex
\section{Conclusion}
\label{sec:conclusion}

In this paper we presented our game video stream analysis toolchain that can be used to extract information about how players interact with games by analyzing video streams from game streaming platforms without having access to internal game data.

We used data extracted by our toolchain to analyze the influence exerted on players by the game mechanics of Fortnite. We identified landing, activity, and kill hot spots which indicate that the game machanics of Fortnite promote certain areas on the map. We also found that the behavior of beginners and experienced players differs, which can be seen by comparing the places where players choose to start the game.

Referring to our observations, we argue that game designers can improve their games by adapting certain game mechanics. For example, our results indicate that beginners landing in hot spots feel lower enjoyment and satisfaction. Game designers could use game mechanics to steer the actions of beginners towards more enjoyable gaming sessions.

Furthermore, our results demonstrate the existence of activity hot spots on the one hand, but also locations with hardly any activity on the other hand. %Here, we can observe that designers of successful games already use this knowledge to mitigate imbalanced map usage. 
Epic Games already uses such information to regularly update the map of Fortnite. Newer versions of the game changed locations where we found little player activity\footnote{https://fortniteskins.net/maps/, last accessed: 2018-10-13}.

The cooperation with game developers could offer opportunities for future work. Applying our findings to current games and evaluating player enjoyment could yield additional insights into the influence of game mechanics on player enjoyment. To allow for easier reproducibility we make the source code, data, and instructions publicly available~\footnote{\label{reprofootnote}https://github.com/FortniteVideoAnalysis/FortniteVideoAnalysis}.